\documentclass{nature}

\usepackage{amsmath}
\usepackage[pdftex]{graphicx}
\usepackage{dcolumn}  
\usepackage{bm}           
\usepackage{dsfont}
\usepackage{color}
\usepackage{braket}
\usepackage{hyperref}

\newcounter{defcounter}
\usepackage{amsthm}

\title{Environmental Control Extends Beyond Quantum Dephasing in Exciton Energy Transfer} 

\author{Junhua Zhou$^{1,*}$, Tianrui Chen$^{1,*}$, Dehao Yuan$^{1}$, Enhu He$^{1}$, Vandana Tiwari$^{2}$, Maxim Gelin$^{3}$, Francoise Remacle$^{4}$, R. J. Dwayne Miller$^{5}$, Fulu Zheng$^{1}$, Ajay Jha$^{6,7}$, Hong-Guang Duan$^{1}$} 

\begin{document} 

\maketitle 

\begin{affiliations} 
\item Department of Physics, School of Physical Science and Technology, Ningbo University, Ningbo, 315211, P.R. China 
\item Linac Coherent Light Source, SLAC National Accelerator Laboratory, Menlo Park, California 94025.
\item School of Science, Hangzhou Dianzi University, Hangzhou, Zhejiang 310018, People’s Republic of China 
\item Theoretical Physical Chemistry, University of Liège, 4000 Liège, Belgium. 
\item The Departments of Chemistry and Physics, University of Toronto, 80 St.\ George Street, Toronto Canada M5S 3H6
\item Rosalind Franklin Institute, Harwell, Oxfordshire OX11 0QX, United Kingdom
\item Department of Pharmacology, University of Oxford, Oxford, OX1 3QT United Kingdom  \\ 

$^*$These authors contributed equally to this work. \\ 
\centerline{\underline{\date{\bf \today}}} 
\end{affiliations} 

\begin{abstract} 

Excitation-energy transfer underpins the conversion of light into usable energy in photosynthetic organisms and serves as a paradigm for evolutionary optimized transport in open quantum systems. Although this process is often described as incoherent thermally assisted hopping, such descriptions become inadequate when electronic coupling, vibronic interactions, and environmental fluctuations occur on comparable energy scales. Determining how the environment controls transport therefore remains a fundamental challenge. Here, we use temperature-dependent two-dimensional electronic spectroscopy to investigate energy transfer in the photosynthetic antenna protein allophycocyanin (APC) over the range 10-296 K. The dominant $\beta \rightarrow \alpha$ transfer step exhibits a pronounced non-monotonic temperature dependence: the transfer time decreases from $\sim$400 fs at 10 K to $\sim$200 fs near 30 - 40 K before increasing again to $\sim$400 fs at 296 K. In contrast, the homogeneous optical dephasing time decreases monotonically across the same temperature range. To interpret these observations, we model APC as a vibronically coupled excitonic dimer interacting with a structured environment and solve the dynamics using hierarchical equations of motion. Conventional fixed-bath models, including Drude-Lorentz and explicit intermolecular-mode spectral densities, fail to reproduce the observed turnover. Quantitative agreement is obtained only when the low-frequency sector of the environmental spectral density is allowed to anharmonically evolve strongly with temperature, while the high-frequency bath remains essentially unchanged. More broadly, these findings demonstrate that transport efficiency is controlled not simply by the magnitude of environmental fluctuations, but by the distribution of environmental spectral weight across frequency space, providing new experimental constraints on theories of molecular transport in complex quantum environments. 

\end{abstract} 


Excitation-energy transfer is one of the most fundamental dynamical processes in nature, governing how absorbed optical energy is redistributed through the molecular antenna before it is converted, stored, or dissipated. The same physical problem arises across a wide range of systems, including photosynthetic light harvesting\cite{Ref1, Ref2}, molecular aggregates\cite{Ref2a}, organic semiconductors\cite{Ref2b}, and emerging excitonic materials\cite{Ref2c}, where electronically excited states evolve under the combined influence of intermolecular interactions and coupling to a fluctuating environment. Among these systems, photosynthetic pigment-protein complexes occupy a unique position. Their chromophore arrangements are defined with near-atomic precision by the protein scaffold, yet the surrounding environment remains sufficiently dynamic to influence excitation transport on timescales comparable to the excited electronic motion itself. As a result, photosynthetic complexes have become model systems for investigating the fundamental principles governing energy flow in complex molecular environments.

Historically, excitation-energy transfer has been understood through theories developed for well-defined limiting regimes. In the weak electronic coupling limit, F{\"o}rster resonance energy transfer (FRET) theory\cite{Ref3, Ref4} describes transport as incoherent hopping between localized chromophores, with transfer rates determined by the electronic coupling and the spectral overlap between donor emission and acceptor absorption. This framework has been remarkably successful in explaining many aspects of molecular energy transfer and continues to serve as the conceptual foundation for interpreting transport in weakly interacting systems. However, many photosynthetic antenna complexes operate outside this regime. The close spatial proximity of pigments often gives rise to electronic couplings comparable to energetic disorder and environmental reorganization energies, leading to excitonic states that are delocalized over multiple chromophores\cite{Ref5, Ref6}. Under such conditions, the assumptions underlying classical hopping theories become progressively less reliable, and transport can no longer be described solely in terms of localized donor and acceptor states. The resulting picture is further complicated by the role of the environment. In conventional treatments, the protein scaffold is frequently regarded as a source of thermal fluctuations that induce relaxation and dephasing of electronic excitations. Increasing experimental and theoretical evidence, however, suggests that the environment plays a more active role. Protein and solvent motions span a broad range of frequencies and timescales, from rapid intramolecular vibrations to slower collective structural fluctuations, and these different environmental degrees of freedom need not influence transport in the same manner. Consequently, energy-transfer efficiency may depend not simply on the overall strength of system-environment interactions, but on how environmental spectral weight is distributed across frequency space. Understanding which sectors of the environment promote, hinder, or regulate excitation transport remains one of the central challenges in the study of photosynthetic energy transfer and, more broadly, in the physics of open quantum systems.

Addressing these questions requires theoretical approaches that move beyond perturbative descriptions of system-environment interactions. Over the past two decades, significant progress has been achieved through the development of open-quantum-system methods capable of treating excitonic dynamics across a broad range of coupling strengths and timescales \cite{Ref7, Ref8, Ref9, Ref10, Nancy SA, Plenio JCP, Renger Biophys J, Mukamel JCP 1998}. Among these, the hierarchical equations of motion (HEOM) \cite{Ref11, Ref12, Ref13, Ref14} and quasi-adiabatic path-integral (QUAPI) approaches \cite{Ref15, Ref16, Ref17, Ref18} provide numerically exact descriptions of energy transfer while fully retaining the influence of environmental fluctuations. These developments have fundamentally changed how photosynthetic transport is viewed. Rather than treating the environment solely as a source of decoherence, modern open-system theories describe energy transfer as arising from the interplay between electronic coupling and a structured bath whose dynamics span multiple frequency scales.

A central ingredient in such descriptions is the environmental spectral density, which specifies how electronic excitations couple to vibrational and structural motions of the surrounding protein matrix. Different forms of the spectral density can give rise to qualitatively different transport behavior, even when the overall strength of system-bath coupling remains similar. In particular, theoretical studies have shown that the relative contributions of low-frequency collective motions and higher-frequency vibrational modes can strongly influence relaxation pathways, coherence lifetimes, and the temperature dependence of energy transfer \cite{Renger SD 2008, Ishizaki and Thorwart}. Despite these advances, direct experimental tests capable of distinguishing between competing environmental models remain relatively rare. Consequently, a central challenge is to identify experimental observables that can reveal not only the magnitude of environmental coupling, but also which spectral weighted components of the environment fluctuations are most relevant for transport.

Temperature-dependent spectroscopy provides a particularly powerful route to addressing this challenge. Because low-frequency environmental modes respond much more strongly to temperature than high-frequency intramolecular vibrations, varying temperature offers a means of selectively probing different regions of the environmental spectrum. Measurements extending into the cryogenic, 10 Kelvin, regime are especially informative because they suppress thermal broadening while retaining the influence of specific environmental fluctuations that are often obscured at ambient conditions. By following energy-transfer dynamics over a wide temperature range, it therefore becomes possible to determine whether transport is governed primarily by conventional dephasing processes with conventional spectral density of states or by more specific interactions with the environment that become separable at lower temperatures. Among the ultrafast techniques required to observe the relevant dynamics, two dimensional electronic spectroscopy (2DES)\cite{Ref6, Ref22, Ref23, Cui PNAS} is uniquely suited to this task. By correlating excitation and detection frequencies with femtosecond temporal resolution, 2DES directly tracks the flow of excitation energy while simultaneously resolving the broadening mechanisms that accompany it. Cross-peaks provide direct signatures of population transfer between spectrally distinct excitonic states, whereas the evolution of spectral linewidths reports on homogeneous and inhomogeneous broadening. Consequently, transport dynamics and dephasing can be quantified within a single experiment, allowing direct tests of whether transfer rates are intrinsically linked to coherence loss or controlled by more subtle environmental effects.

The allophycocyanin (APC) antenna complex provides an ideal model system for such investigations. APC is a phycobiliprotein antenna complex from cyanobacterial phycobilisomes and consists of a trimer of identical $\alpha$$\beta$ heterodimers \cite{Ref19, Ref20, Ref21}. Each monomer contains only two covalently bound chromophores, one phycocyanobilin pigment in the $\alpha$-subunit and one in the $\beta$-subunit, held at a fixed separation of approximately 21 \AA. This well-defined geometry leads to strong electronic coupling and an ultrafast energy-transfer step from the higher-energy $\beta$ pigment to the lower-energy $\alpha$ pigment on a sub-picosecond timescale. The presence of a dominant and spectroscopically resolvable transfer pathway, without the multichromophore congestion typical of larger antenna complexes, allows APC to isolate how environmental dynamics influence energy transfer with minimal ambiguity. Building on our recent temperature-dependent investigation of coherence in APC, which established the ultrafast loss of electronic coherence and the predominance of vibrational coherences, the present work addresses a complementary question: how the temperature evolution of the environment governs excitation-energy transfer itself \cite{APCI tianrui}. Here, we combine previously reported 2DES measurements at 10, 80 and 296 K with newly acquired intermediate-temperature data with numerically exact open-quantum-system simulations based on the hierarchical equations of motion to establish the complete temperature dependence of excitation-energy transfer in APC. We find that the $\beta \rightarrow \alpha$ transfer time exhibits a pronounced turnover, accelerating from approximately 400 fs at 10 K to about 200 fs near 30-40 K before slowing again at higher temperatures. In contrast, the homogeneous dephasing time evolves monotonically across the same range. This observation demonstrates that transfer efficiency cannot be understood through a single dephasing metric. Instead, our results show that temperature-dependent coupling to low-frequency environmental modes plays a distinct and active role in regulating ultrafast energy transport, identifying the low-frequency sector of the environment as a key control parameter in photosynthetic exciton dynamics.

\section*{Results} 

\begin{figure}[h!]
\begin{center}
\includegraphics[width=14.0cm]{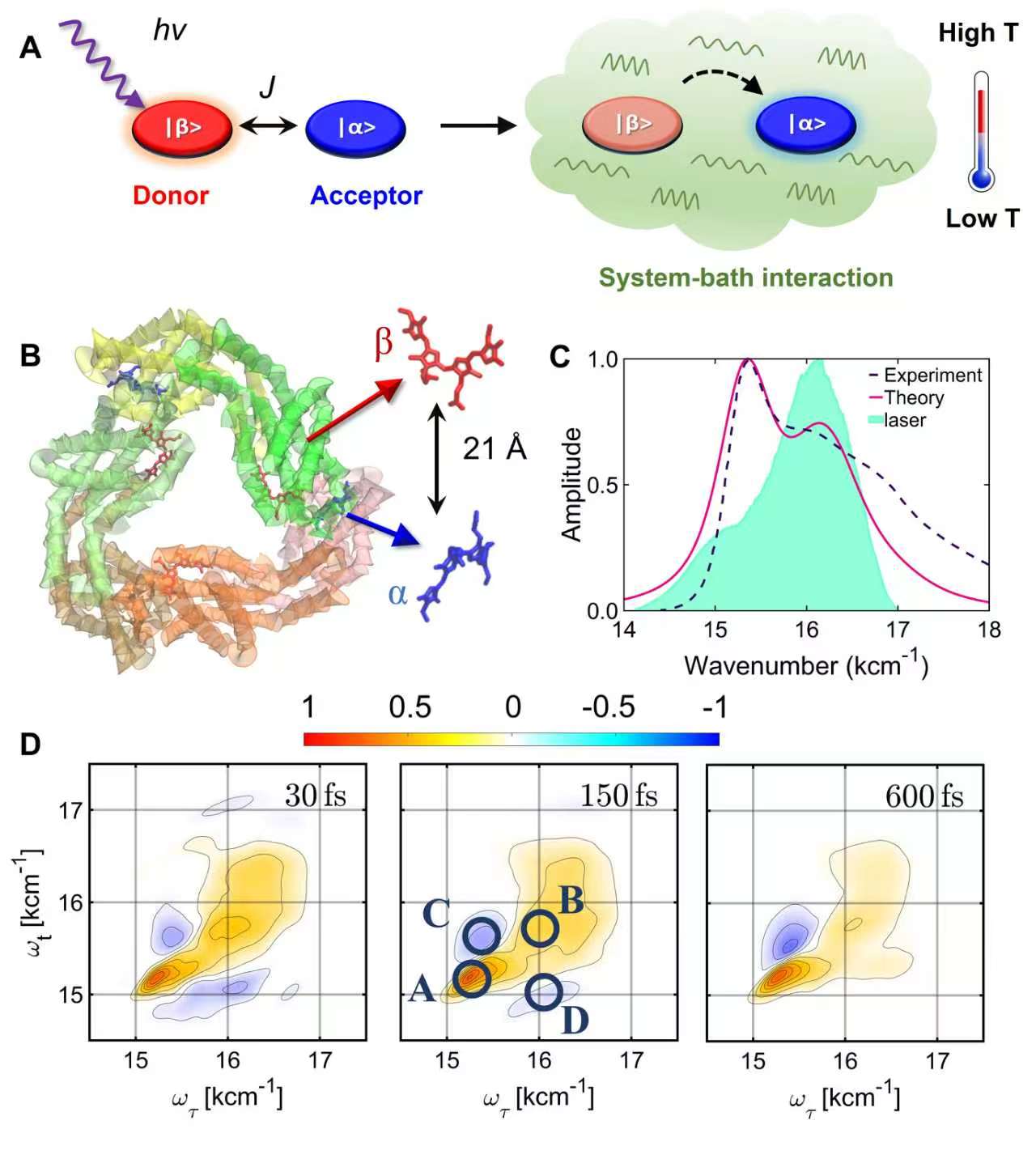}
\caption{\label{fig:Fig1}APC structure and temperature-dependent 2DES. (A) Conceptual illustration of excitation-energy transfer between donor and acceptor embedded within a fluctuating environment. (B) Crystal structure of the APC trimer highlighting the strongly coupled $\alpha$ and $\beta$ pigments. (C) Experimental (black dashed) and calculated (red) absorption spectra together with the excitation laser spectrum (blue). (D) Representative absorptive 2DES spectra recorded at 10 K for selected waiting times, showing ground-state bleaching/stimulated emission (positive) and excited-state absorption (negative) features.} 
\end{center}
\end{figure}

The schematic picture of excitonic energy transport between donor and acceptor has been illustrated in Fig.\ \ref{fig:Fig1} A. The photosynthetic energy transfer has been embedded by noisy environment. The structure of the APC trimer is shown in Fig.\ \ref{fig:Fig1} B. Each monomer consists of an $\alpha\beta$ heterodimer containing two phycocyanobilin chromophores. The pigments associated with the $\alpha$- and $\beta$-subunits are highlighted in blue and red, respectively. Within each heterodimer, the two pigments are separated by $\sim$21 \AA, giving rise to a relatively strong excitonic coupling of $\sim$160 cm$^{-1}$. In contrast, couplings between pigments belonging to different monomers are substantially weaker owing to their larger spatial separation. Details of the electronic-structure calculations and excitonic-coupling analysis are provided in the Theoretical Calculations section and the Supporting Information (SI). The experimental absorption spectrum of APC and the calculated absorption spectrum are compared in Fig.\ \ref{fig:Fig1} C, shown as black dashed and red solid lines, respectively. The spectral bandwidth of the excitation pulses used in the 2DES experiments is overlaid as the blue shaded region. The good agreement between experiment and theory confirms that the adopted excitonic model captures the essential spectral properties of the complex. Details of sample preparation and experimental conditions are provided in the Materials and Methods section.

\subsection{Temperature-dependent two-dimensional electronic spectra} 

Temperature-dependent 2DES measurements were performed using our home-built spectrometer over the temperature range 10-296 K. We begin by considering the spectra recorded at 10 K, where thermal broadening is minimized and the excitonic features are most clearly resolved. Representative real-valued absorptive 2DES spectra measured at waiting times of 30, 150 and 600 fs are shown in Fig.\ \ref{fig:Fig1} D. Positive spectral features correspond to ground-state bleaching (GSB) and stimulated emission (SE), whereas negative features arise from excited-state absorption (ESA). At the earliest waiting time (T = 30 fs), two dominant positive peaks, labeled A and B, are clearly resolved. Both peaks exhibit pronounced elongation along the diagonal direction, indicative of substantial inhomogeneous broadening within the ensemble. Additional higher-energy spectral features are also observed and are assigned to vibronic progression of the excitonic transitions. The negative features labeled C and D correspond to ESA contributions associated with the excited-state manifold of APC. As the waiting time increases to 60 fs, the diagonal elongation of the main excitonic peaks becomes noticeably reduced, accompanied by a narrowing of the overall spectral lineshape. Beyond this timescale, the spectral profile evolves more gradually, and the spectra recorded at 150, 300 and 600 fs retain the same overall structure, with well-resolved diagonal and cross-peak features throughout the experimental time window. 

To examine the influence of temperature, analogous 2DES measurements were performed at 20, 30, 40, 50, 70, 150 and 296 K. The complete dataset is presented in the SI. The most prominent temperature-dependent change is a progressive broadening of the anti-diagonal linewidth with increasing temperature, reflecting enhanced homogeneous broadening and faster optical dephasing. Nevertheless, the energy separation between the two excitonic states remains sufficiently large that both diagonal peaks and the associated cross-peaks remain spectroscopically distinguishable even at 296 K. This clear spectral resolution enables direct tracking of the energy-transfer dynamics over the full temperature range investigated in this work.

\subsection{Energy transfers between two cofactors in APC complex} 

\begin{figure}[h!]
\begin{center}
\includegraphics[width=12.0cm]{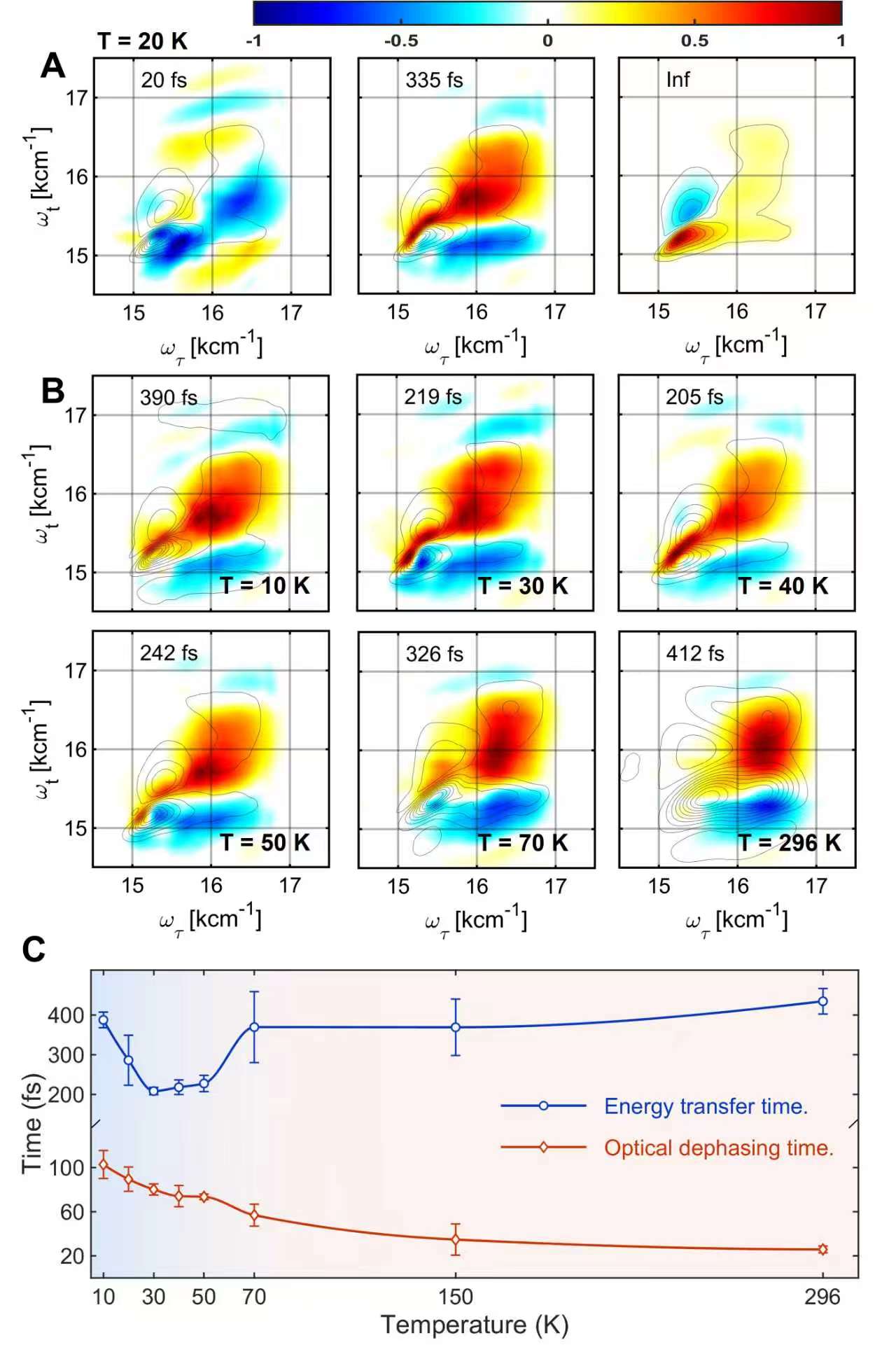}
\caption{\label{fig:Fig2}Temperature dependence of excitation-energy transfer. (A) Representative 2DDAS obtained from global analysis at 20 K. (B) Transfer-associated 2DDAS at different temperatures showing the persistence of the same energy-transfer pathway. (C) Extracted energy-transfer times (blue) and homogeneous optical dephasing times (red) as a function of temperature. Transfer exhibits a pronounced optimum near 30-40 K, whereas dephasing decreases monotonically. } 
\end{center}
\end{figure}

To quantify the energy-transfer dynamics, we performed global analysis of the temperature-dependent 2DES datasets. The measured spectra were assembled into 3D data cubes and analysed using a multiexponential fitting procedure. Details of the fitting algorithm and validation tests are provided in the SI. The resulting two-dimensional decay-associated spectra (2DDAS) obtained at 20 K are shown in Fig.\ \ref{fig:Fig2} A. The fastest component exhibits a characteristic timescale of $\sim$20 fs and is dominated by spectral broadening and lineshape evolution occurring immediately after photoexcitation. A second component with a lifetime of 335 fs displays positive amplitudes at the higher-energy diagonal peak and corresponding negative amplitudes at lower energy. This spectral pattern is characteristic of population transfer from the higher-energy excitonic state to the lower-energy state and therefore reports directly on the dominant energy-transfer process within the APC dimer. A long-lived component accounts for the remaining population that persists beyond the experimental time window. The same analysis was applied to all temperatures investigated in this work. Representative 2DDAS obtained at 10, 30, 40, 50, 70 and 296 K are shown in Fig.\ \ref{fig:Fig2} B. In each case, the transfer-associated component exhibits the same characteristic spectral signature, confirming that the dominant relaxation pathway remains unchanged across the entire temperature range. The extracted transfer times are 390 fs, 219 fs, 205 fs, 242 fs, 326 fs and 412 fs for temperatures of 10, 30, 40, 50, 70 and 296 K, respectively. 

The temperature dependence of the transfer time is summarized in Fig.\ \ref{fig:Fig2} C. Rather than varying monotonically, the transfer exhibits a pronounced minimum between $\sim$30 and 50 K, corresponding to the fastest energy-transfer dynamics observed in the experiment. Transfer becomes slower both at lower and at higher temperatures. The reproducibility of this trend is reflected in the experimental uncertainties shown as error bars. To compare the transfer dynamics with homogeneous optical dephasing, we also analysed the anti-diagonal linewidth of the lowest-energy excitonic peak. The corresponding homogeneous optical-dephasing times are plotted as a red line in Fig.\ \ref{fig:Fig2} C. In contrast to the transfer dynamics, the optical-dephasing time decreases monotonically with increasing temperature, indicating progressively stronger homogeneous broadening at elevated temperatures. The distinct temperature dependences of transfer and optical dephasing reveal that the two observables are not directly correlated over the full temperature range investigated.

\subsection{Modeling and theoretical calculations} 

\begin{figure}[h!]
\begin{center}
\includegraphics[width=12.0cm]{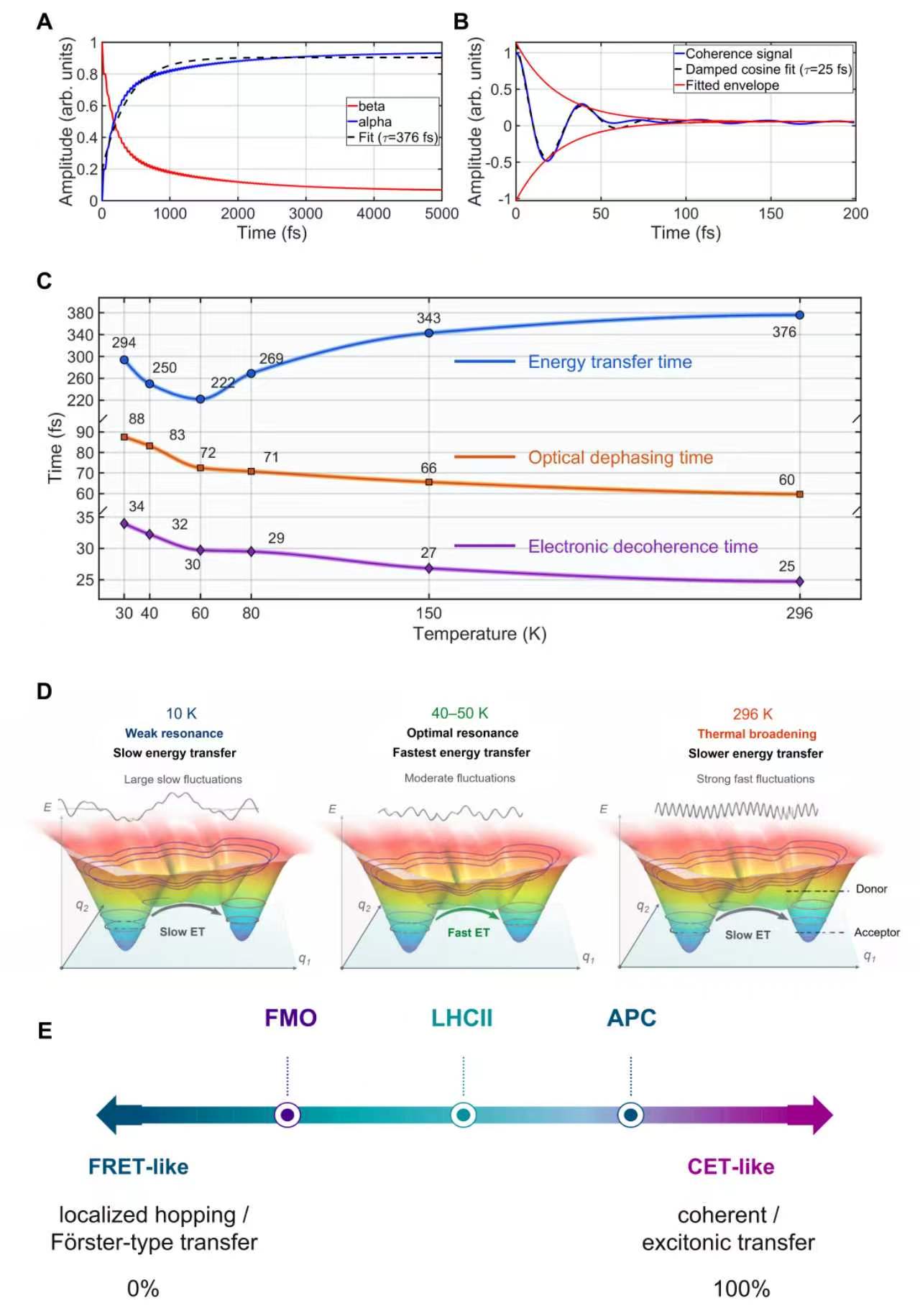}
\caption{\label{fig:Fig3}HEOM simulations identify the environmental origin of the transport turnover. (A) Representative simulated population dynamics and exponential fit. (B) Simulated electronic decoherence dynamics. (C) Calculated temperature dependences of energy-transfer, optical-dephasing, and electronic-decoherence times. (D) Schematic illustrating the influence of low-frequency environmental modes on the excitonic energy landscape. (E) Conceptual placement of APC within the continuum from Förster-like hopping to coherent excitonic transport. } 
\end{center}
\end{figure}

To gain microscopic insight into the experimentally observed temperature dependence of energy transfer, we constructed a system-bath model of the APC dimer based on {\em ab initio} electronic-structure calculations. The site energies of the $\alpha$ and $\beta$ pigments, their electronic coupling, and the relevant excited-state vibrational modes were obtained from quantum-chemical calculations. These parameters were subsequently used to construct an exciton-vibrational Hamiltonian describing the coupled electronic and environmental degrees of freedom. The resulting dynamics were simulated using the HEOM formalism \cite{Ref11, HEOM2}, which provides a numerically exact treatment of system-bath interactions over a wide range of coupling strengths and temperatures. Details of the electronic-structure calculations, model construction, and HEOM implementation are provided in the Materials and Methods section and in the SI. The calculated population dynamics for all temperatures are presented in Sec.\ VI of the SI. To facilitate comparison with experiment, the simulated population-transfer dynamics were fitted using the same exponential analysis employed for the experimental data. An example of the calculated population dynamics and the corresponding fit at 296 K is shown in Fig.\ \ref{fig:Fig3} A. The extracted transfer time is 376 fs at 296 K, in excellent agreement with the experimentally determined value. Using the same framework, we also calculated the electronic decoherence dynamics between the two excited states. Representative results are shown in Fig.\ \ref{fig:Fig3} B, yielding an electronic decoherence time of approximately 25 fs at 296 K. The complete set of calculated population and electronic decoherence dynamics is provided in the SI. The temperature dependence of the calculated transfer times is summarized in Fig.\ \ref{fig:Fig3} C. Remarkably, the simulations reproduce the experimentally observed non-monotonic behavior, with the fastest energy-transfer dynamics occurring between approximately 30 and 60 K. In contrast, the calculated optical-dephasing times shown in Fig.\ \ref{fig:Fig3} C decrease monotonically with increasing temperature, consistent with the trend extracted from the anti-diagonal linewidth analysis of the 2DES measurements. The electronic decoherence times exhibit a similar monotonic decrease. The ability of the model to simultaneously reproduce the transfer turnover and the monotonic optical-dephasing behavior provides a stringent test of the underlying microscopic description. Figure 3D provides a physical interpretation of these contrasting trends. While the optical-dephasing time decreases monotonically with temperature owing to enhanced high-frequency thermal fluctuations, the low-frequency environmental coordinates evolve non-monotonically, transiently improving donor-acceptor resonance near 30-40 K before thermal broadening ultimately slows excitation-energy transfer.

To investigate the origin of the transport turnover, we examined different descriptions of the environmental coupling within the HEOM framework. As additional tests, we explicitly included intermolecular (Peierls-type) vibrations in the vibronic dimer model \cite{Thomas Renger paper, ZhaoYang1} (SI, Sec. VI(3)). These modes produced only a weak modulation of the population-transfer dynamics, consistent with their relatively small vibronic coupling strengths. We also considered a super-Ohmic spectral density \cite{Yang2012,Ritschel2014}, which captures the overall magnitude and broad temperature dependence of the transfer dynamics but does not quantitatively reproduce the experimentally observed turnover. Together, these tests show that conventional fixed-bath descriptions, including explicit intermolecular modes and a super-Ohmic bath, are insufficient to account for the full temperature dependence of the experiment. Previous theoretical studies have suggested that low-frequency environmental fluctuations and quasi-static structural disorder can play an important role in determining excitation transport in photosynthetic systems \cite{Darius Abramavicius Ref}. Motivated by these studies, we partitioned the environmental spectral density into low- and high-frequency components. The high-frequency contribution describes the conventional vibrational environment arising from pigment vibrations, protein motions, and solvent fluctuations. In contrast, the low-frequency contribution captures slowly varying environmental coordinates that modulate the excitonic energy landscape on timescales comparable to the transport dynamics. Calculations employing temperature-independent spectral densities were unable to reproduce the experimentally observed turnover in transfer time while simultaneously yielding the monotonic dephasing dynamics. Agreement with experiment was obtained only when the low-frequency sector of the spectral density was allowed to evolve with temperature. As shown in Fig.\ \ref{fig:Fig3} C, the resulting model reproduces both the optimum in energy-transfer efficiency near 30-40 K and the monotonic reductions of optical-dephasing and electronic-decoherence times with increasing temperature. These results indicate that the low-frequency environmental modes provide an essential contribution to the observed transport behaviour and motivate a more detailed discussion of their physical role in the following section.

\section*{Environmental Control of Excitation Energy Transfer} 

The central result of this work is that excitation-energy transfer and homogeneous optical dephasing exhibit fundamentally different temperature dependences. The experimentally observed $\beta\rightarrow\alpha$ transfer time displays a pronounced minimum near 30-40 K, whereas the homogeneous dephasing time decreases monotonically over the same temperature range. This simple observation immediately demonstrates that transport cannot be inferred from homogeneous dephasing alone. If the transfer rate were determined solely by dephasing, k$_{ET}$(T) = f[$\tau_{hom}$(T)], where f is a monotonic function, the measured monotonic decrease of $\tau_{hom}$ would necessarily produce a monotonic variation of k$_{ET}$. Since this is not observed experimentally, the APC data require an additional temperature-dependent environmental coordinate that independently influences transport. The HEOM calculations identify this coordinate as the low-frequency sector of the environmental spectral density. In our model 
\begin{equation} 
\label{eq:spectral density 2} 
J_i (\omega, T) = J_{(i,low)} (\omega, T) + J_{(i,high)} (\omega), 
\end{equation} 
with only the low-frequency reorganization energies $\lambda_{(i,low)}$(T) allowed to vary with temperature. This means that the total derivative of the transfer rate can be decomposed as 
\begin{equation} 
\label{eq:formula 1} 
\frac{dk_{ET}}{dT} = ( \frac{\partial k_{ET}}{\partial T} )_{\lambda _{low}} + (\frac{\partial k_{ET}}{\partial \lambda_{low}})_{T} \frac{d\lambda_{low}}{dT}. 
\end{equation} 
Conventional fixed-bath descriptions retain only the first contribution and therefore predict a monotonic temperature dependence. The experimentally observed turnover is reproduced only when the second contribution arising from temperature-dependent low-frequency environmental coupling is included. Importantly, the same model simultaneously reproduces the monotonic dephasing behaviour (Fig.\ \ref{fig:Fig3}D), providing strong evidence that the transport optimum originates from changes in the low-frequency environment rather than homogeneous dephasing itself. Fig.\ \ref{fig:Fig3}D illustrates the physical mechanism emerging from the combined experiment and theory. At cryogenic temperatures, excitonic states remain strongly coupled to specific low-frequency environmental modes whose characteristic frequencies are comparable to the transfer timescale. These slow fluctuations continuously modulate the donor-acceptor energy gap and reduce transport efficiency. As temperature initially increases, coupling to these low-frequency modes weakens, allowing energy transfer to accelerate. Beyond approximately 30-40 K, thermal fluctuations dominate, increasing homogeneous broadening and reducing spectral overlap, causing transport to slow again. The transfer optimum therefore reflects the competition between temperature-dependent coupling to low-frequency environmental modes and thermally activated dephasing.

This interpretation also places the present results in the broader context of static disorder. Rather than representing a purely temperature-independent distribution of site energies, slowly varying environmental coordinates evolve according to their own thermal response and actively regulate transport. In the limit where the representative low-frequency mode approaches zero frequency, the present description reduces to the static-disorder spectral density proposed by Shi and co-workers \cite{Shi2024}.

The broader implication of this work is illustrated schematically in Fig.\ \ref{fig:Fig3}E. APC occupies an intermediate transport regime between incoherent Förster hopping and fully delocalized excitonic transport, where neither electronic coupling nor environmental interactions can be regarded as small perturbations. Our results show that, in this regime, the relevant environmental control variable is not simply the overall magnitude of system-bath coupling, but the distribution of environmental spectral weight across frequency space. Consequently, two environments with comparable total reorganization energies can produce qualitatively different transport dynamics because low-frequency environmental modes modify the excitonic energy landscape in a fundamentally different manner from faster fluctuations, which primarily contribute to thermal relaxation and homogeneous dephasing. Temperature-dependent 2DES therefore provides considerably more than a measurement of transfer kinetics; it serves as a direct spectroscopic probe of the environmental degrees of freedom governing molecular energy transport. By combining experiment with non-perturbative HEOM simulations, the present work identifies the specific sector of the environmental spectrum responsible for the observed transport behaviour and demonstrates how temperature-dependent coupling to low-frequency modes produces dynamics that cannot be captured by conventional fixed-bath descriptions. These findings therefore establish a direct experimental framework for testing microscopic models of structured environments in molecular open quantum systems.

The scientifically justified connection to the `quantum world' should remain equally precise. These results do not prove that biology exploits long-lived electronic coherence for function, nor do they imply that APC realizes a quantum algorithm. They do prove something narrower and more robust: in a real molecular dimer, transport can be optimized or suppressed by engineering the low-frequency environment, while homogeneous dephasing alone is an insufficient design metric. That conclusion is directly relevant to present-day quantum and excitonic platforms, where structured noise and bath engineering already serve as controllable resources. APC therefore offers an experimentally grounded molecular example of a general design principle: the environment is not only a sink of coherence, but also a spectral object that can steer transport when its low-frequency sector is properly controlled.

\section*{Conclusion} 

Excitation-energy transfer is generally viewed as a balance between electronic coupling and environmentally induced dephasing. The present work demonstrates that this description is incomplete. By combining cryogenic 2DES with quantitatively accurate HEOM simulations, we show that the temperature dependence of energy transfer in the APC antenna cannot be explained by homogeneous dephasing alone. Instead, the experimentally observed transport optimum originates from temperature-dependent coupling to the low-frequency sector of the environmental spectral density, while the high-frequency environment predominantly governs optical dephasing. These results identify a distinct environmental control parameter for excitation transport. Rather than being determined solely by the overall strength of system-bath interactions, transport depends critically on how environmental fluctuations are distributed across frequency space. The work therefore shifts the focus from treating the environment as a passive source of decoherence to viewing it as a structured dynamical system whose low-frequency degrees of freedom actively regulate energy flow. Beyond the specific case of APC, our findings establish a broader framework for investigating transport in molecular open quantum systems that operate in the intermediate regime between incoherent hopping and fully delocalized excitonic dynamics. While the present results do not imply that biological systems exploit long-lived electronic coherence for function, they demonstrate that the spectral structure of the environment provides an independent and experimentally accessible handle for controlling excitation transport. More generally, our results establish cryogenic multidimensional spectroscopy as a quantitative probe of environmental spectral structure, providing a direct experimental route to identifying and ultimately engineering the environmental degrees of freedom that control energy transport in complex molecular quantum systems.
%

\section*{Materials and Methods} 

\subsection{Sample preparation.} 

The cross-linked allophycocyanin trimer complex (APC) was ordered from Crystal Biotech company (website: http://www.crystalbiosystems.com/productdetail-471show.html) and we used the sample without further modifications. The APC complex was initially dissolved in the  solution (20 mM PB, 2 mM EDTA) and diluted to the maximum of OD 0.25 (absorption spectrum) at wavelength of 650 nm. For the 296 K measurement, we moved the sample into a quartz cuvette with sample thickness of 0.5 mm. The cuvette has been mounted in a home-built shaker system with X and Y moving axes. The shaking speed can be modulated and well controlled by power supply. Also, a home-built XYZ-delay-stage system has been developed to modulate the laser spot focusing on sample. By this, we could optimize the sample position and find the best beam size and coordinates for measuring of TA, TG and 2DES. The best signal-to-noise ration can be achieved by reducing scattering. We mixed APC sample with glycerol with ratio (APC sample v1/glycerol v2 = 2:8). They were well mixed and put in the home-built cell system. This device has been put in the cryostat (ARS system with compressed Helium) to cool sample to 10 K and other temperatures. The detailed measuring conditions of laser system and spectroscopic setups, NOPA and TG spectrometer are described in the next section. 

\subsection{2D and TG Electronic measurements with experimental conditions.}

The details of the experimental setup follow previous reports from our group \cite{Jing JACS, Tiwari JACS}. In brief, two-dimensional electronic spectra were recorded using a diffractive optics-based, all-reflective 2D spectrometer providing phase stability of $\lambda/160$. Excitation pulses were generated by a home-built nonlinear optical parametric amplifier (NOPA), pumped by a commercial femtosecond laser system (Spectra Physics, Newport). The pulses were compressed to $\sim$12 fs using a combination of a deformable mirror (OKO Technologies, 19-channel) and a pair of chirp mirrors (Munich Ultrafast Innovations GmbH). Their temporal profile was characterized by frequency-resolved optical gating (FROG), with the traces analyzed using the commercial package FROG3 (Femtosecond Technologies). The resulting broadband spectrum had a $\sim$100 nm full width at half maximum (FWHM), centered at 625 nm, sufficient to cover the main electronic transitions into the lowest excited states. For 2DES measurements, three beams were focused onto the sample with a spot size of $\sim$130~$\mu$m, generating a photon-echo signal along the phase-matching direction. The emitted signal was collected using a Sciencetech 9055F spectrometer coupled to a CCD linear array detector (Entwicklungsb{\"u}ro Stresing). Transient grating (TG) spectra were recorded at multiple waiting times $T$ by scanning the population delay from -200 fs to 3 ps with 3 fs steps. At each delay point, 200 individual spectra were averaged to improve the signal-to-noise ratio. For all experiments, the excitation pulse energy was attenuated to $\sim$10 nJ at a 1 kHz repetition rate. Phasing of the TG data was achieved using the ``invariant theorem” procedure described in Ref.\ \citeonline{David2001}. The 2DES datasets recorded at 10 K, 80 K and 296 K have been reported previously in Ref. \cite{APCI tianrui}. In the present work, these measurements are combined with newly acquired intermediate-temperature datasets (20-150 K) to establish the complete temperature dependence of excitation-energy transfer.

\subsection{Model Hamiltonian and parameters.} 

For an open quantum system, the total Hamiltonian can be generally expressed as 
\begin{equation}
\label{eq:Htot}
H = H_S + H_B + H_{SB}, 
\end{equation}
where $H_S$ denotes the Hamiltonian of the system of interest, $H_B$ represents the Hamiltonian of the surrounding environment (bath), and $H_{SB}$ captures the interaction between the system and the bath. To model excitation energy transfer in the APC light-harvesting complex, we consider an effective vibronic Hamiltonian for the strongly coupled $\beta_{84}$--$\alpha_{84}$ pigment pair. In this two-site representation, $\ket{1}$ denotes the localized excitation on $\beta_{84}$ and $\ket{2}$ denotes the localized excitation on $\alpha_{84}$. The system Hamiltonian can be written as 
\begin{equation}
\label{eq:Hs}
H_S = \sum_{m=1}^{2} \ket{m} \left( \epsilon_m+h_m \right) \bra{m} + J \left( \ket{1}\bra{2} + \ket{2}\bra{1} \right),
\end{equation}
where $\epsilon_1=16060~\mathrm{cm}^{-1}$ and $\epsilon_2=15300~\mathrm{cm}^{-1}$ are the site energies of the $\beta_{84}$ and $\alpha_{84}$ pigments, respectively. The electronic coupling is $J=-157~\mathrm{cm}^{-1}$, which facilitates population transfer and electronic coherence between the two pigments. The vibrational structure of each site is included through the site-local vibrational Hamiltonian 
\begin{equation}
\label{eq:h}
h_m = \hbar\Omega \left( a^\dagger a+\frac{1}{2} \right) + \kappa_m Q, 
\end{equation}
where $\Omega=800~\mathrm{cm}^{-1}$ is the frequency of the intramolecular vibrational mode, and $a^\dagger$ ($a$) are the bosonic creation (annihilation) operators. $Q$ is the vibrational reaction coordinate, and $\kappa_m$ is the site-dependent vibronic coupling strength. This term modulates the local excitation energy as a function of nuclear motion and enables electronic-vibrational state mixing. The bath Hamiltonian is described as a set of harmonic oscillators, 
\begin{equation}
\label{eq:HB}
H_B = \sum_{\xi} \left( \frac{p_{\xi}^{2}}{2m_{\xi}} + \frac{1}{2}m_{\xi}\omega_{\xi}^{2}x_{\xi}^{2} \right), 
\end{equation}
where $\xi$ labels the bath modes. The system-bath interaction is assumed to be diagonal in the localized electronic basis, 
\begin{equation}
\label{eq:HSB}
H_{SB} = \sum_{m=1}^{2} \ket{m}\bra{m}\,\Phi_m, 
\end{equation}
where $\Phi_m$ is the collective bath coordinate coupled to pigment $m$. In the present work, each pigment is coupled to two independent Drude-Lorentz bath components, corresponding to a low-frequency environmental contribution and a high-frequency background contribution. The spectral density of pigment $m$ is therefore written as 
\begin{equation}
\label{eq:Jtotal}
J_m(\omega,T) = J_{m,\mathrm{low}}(\omega,T) + J_{m,\mathrm{high}}(\omega),
\end{equation}
with 
\begin{equation}
\label{eq:Jdrude}
J_{m,r}(\omega,T) = \frac{2\lambda_{m,r}(T)\gamma_r\omega} {\omega^2+\gamma_r^2}, 
\qquad r=\mathrm{low},\mathrm{high}.
\end{equation}
Here, $\lambda_{m,r}$ is the reorganization energy and $\gamma_r$ is the bath relaxation rate. The low-frequency component represents slow environmental fluctuations, whereas the high-frequency component describes faster bath motions. In the temperature-dependent low-frequency Drude--Lorentz model, the temperature dependence is introduced through the low-frequency reorganization energies $\lambda_{m,\mathrm{low}}(T)$, while the high-frequency bath parameters are kept unchanged. The detailed bath parameters used in the simulations are provided in the Supporting Information.

%
\begin{addendum} 
\item This work was supported by the National Key Research and Development Program of China (Grant No.\ 2024YFA1409800), NSFC Grant with No.\ 12274247 and No.\ 12404310, Yongjiang talents program with No.\ 2022A-094-G and 2023A-158-G, Ningbo International Science and Technology Cooperation with No.\ 2023H009, the foundation of national excellent young scientist. 

\item[Supporting information] The details of Hierarchy equation of motion, global fitting approach, detailed data treatment and refinement, 2DES measured at different temperature, modeling and parameters are described. 

\item[Competing Interests] The authors declare that they have no competing financial interests. 

\item[Author contributions] H.G.D. and A.J. conceptualize the project. J.Z. performed the 2DES measurement at different temperatures with the help of T.C., D.Y., E.H. and Z.H.. T.C. and F.Z. performed the {\em ab-initio} calculations and also the 2DES calculations by HEOM with the inputs from M.G. and F.R.. H.G.D., A.J., F.Z. and R.J.D.M. supervised the project. H.G.D. and A.J. did the data interpretation with inputs from R.J.D.M. and V.T.. H.G.D. and A.J. wrote the initial draft of the manuscript. All the authors discussed and refined the final version of manuscript. 

\item[Correspondence] Correspondence of paper should be addressed to R.J.D.M. ~(dwayne.miller@utoronto.ca), F. L. Z. ~(Zhengfulu@nbu.edu.cn), A.J. ~(Ajay.Jha@rfi.ac.uk) and H.-G.D. ~(email: duanhongguang@nbu.edu.cn). 

\end{addendum}
%
\newpage

\newpage

\newpage

\end{document}